\documentclass[english,a4paper]{article}
\usepackage[margin=2cm]{geometry}
\usepackage[T1]{fontenc}
\usepackage[latin9]{inputenc}
\usepackage{amsmath}
\usepackage{amssymb}
\usepackage[numbers]{natbib}
\usepackage{authblk}

\newcommand{\TYNDALL}{Tyndall National Institute, University College Cork,  Ireland} 
\newcommand{\MATHUCC}{School of Mathematical Sciences, University College Cork, Ireland}

\usepackage{babel}

\begin{document}

\title{An odd-number limitation of extended time-delayed feedback control
in autonomous systems}

\author[1,2]{Andreas Amann}
\author[1]{Edward W. Hooton}
\affil[1]{\MATHUCC}
\affil[2]{\TYNDALL}

\maketitle

\abstract{We propose a necessary condition for the successful stabilisation
of a periodic orbit using the extended version of time-delayed feedback
control. This condition depends on the number of real Floquet multipliers
larger than unity and is therefore related to the well-known odd-number
limitation in non-autonomous systems. We show that the period of the
orbit which is induced by mismatching the delay-time of the control
scheme and the period of the uncontrolled orbit plays an important
role in the formulation of the odd-number limitation in the autonomous
case.}



\section{Introduction}

By their nature, chaotic systems are extremely sensitive to external
perturbations, which makes it difficult to predict their future evolution.
However, this sensitivity also allows for the surprising possibility
of \emph{controlling }a chaotic system. As first demonstrated by Ott,
Grebogi and Yorke \citep{OTT90}, an external perturbation can be
grafted in such a way that one of the unstable periodic orbits of
the chaotic attractor becomes stable. This discovery has triggered
a large research activity centred around the oxymoronic term \emph{chaos
control }\citep{SCH07d,BOC00}.

A simple but highly efficient scheme of chaos control was introduced
by Pyragas in \citep{PYR92}. In order to stabilise a periodic orbit
of period $\tau$ the original system is converted into a system of
time-delayed differential equations by adding terms which involve
the difference $x\left(t-\tau\right)-x\left(t\right)$. Here $x\left(t\right)$
denotes a point in the phase space of the original system. Control
terms of this form vanish, whenever a periodic orbit with period $\tau$
is reached and therefore the Pyragas control scheme is automatically
\emph{non-invasive.} The time-delayed feedback control was extended
by Socolar et al. \citep{SOC94} by employing additional control terms
of the form $x\left(t-k\tau\right)-x\left(t-\left(k-1\right)\tau\right)$
for integer values $k>1$. Again, terms of this form vanish on a periodic
orbit of period $\tau$, and therefore enable \emph{non-invasive }control.
This \emph{extended time-delayed feedback control }(EDFC) scheme is
important for practical applications, because it can significantly
increase the range of periodic orbits, which can be stabilised \citep{BEC02}. 

Unfortunately, not all periodic orbits can be stabilised by time-delayed
feedback control, which severely limits its applicability. In particular,
in a non-autonomous system a hyperbolic periodic orbit with an odd
number of Floquet multipliers larger than unity can never be stabilised
by time-delay feedback control. This became known as the \emph{odd-number
limitation} and was proved in \citep{NAK97} for the original Pyragas
control scheme and in \citep{NAK98} for a fairly general class of
EDFCs. While the proofs concerning non-autonomous systems presented
in \citep{NAK97,NAK98} are correct, the autonomous case was not treated
correctly. In particular footnote 2 in \citep{NAK97} claims that
all results regarding the odd-number limitation can be proved for
the autonomous case ``with a slight revision''. More importantly,
Theorem 2 in \citep{NAK98} claims explicitly that the odd number
limitation applies to the autonomous case. However, the proof of this
theorem contains an error in the expansion of a determinant and is
therefore wrong. The non-autonomous case was also discussed in \citep{JUS97}
and it was stated that the Pyragas method for non-autonomous systems
should only be able to stabilise orbits with ``finite torsion'',
but not orbits with a single Floquet multiplier larger then unity.
Although it was not explicitly claimed in \citep{JUS97} that this
statement can be extended to the autonomous case, the casual reader
might not have noticed this subtle point. Based on \citep{NAK97,NAK98,JUS97}
there was therefore a general belief among members of the chaos-control
community that the odd-number limitation also holds in the autonomous
case. 

This changed with the work by Fiedler et al. \citep{FIE07}, who gave
an example of a two-dimensional autonomous system, with precisely
one Floquet multiplier larger than one, which can be stabilised using
the original Pyragas control scheme. This immediately showed that
the odd-number limitation does not hold for autonomous systems and
opened up the exciting possibility that the time-delayed feedback
control could be far more powerful than previously thought. 

While \citep{FIE07} provides a counter example for the original odd-number
limitation, it raises the question, if the odd-number limitation is
outright wrong, or if it holds at least under certain additional conditions.
After all, numerical and experimental evidence never showed any problem
with the odd-number limitation before the publication of \citep{FIE07},
which suggests that for many systems or control schemes the odd-number
limitation might indeed be true. Motivated by this possibility we
showed in our recent work \citep{HOO12} that a modified version of
the odd-number limitation holds for Pyragas type control. Like the
original odd-number limitation, our modified version also involves
the number of Floquet multipliers greater than unity, but in addition
also depends on an analytical expression which involves an integral
of the control force along the desired periodic orbit. Interestingly,
this analytical expression can also be obtained by studying the period
of the orbit which is induced if the system is forced with a delayed
feedback term where the delay time does not match the period of the
orbit in the unforced system. Our modified version of the odd-number
limitation correctly predicts the stability boundaries of the previous
counter example presented in \citep{FIE07}.

In the current work we generalise our previous results on the odd-number
limitation of Pyragas control to the case of EDFC. While we closely
follow the arguments laid out in \citep{HOO12} we intend to keep
the presentation self-contained. The remainder of the paper is organised
as follows: in Section \ref{sec:Statement-of-the} we introduce the
notation and state the main theorem which is then proved in Section
\ref{sec:Proof-of-the}. In the final Section \ref{sec:Discussion-and-conclusion}
we discuss the significance and practical implication of our results.

\section{Statement of the theorem\label{sec:Statement-of-the}}

We consider a dynamical system of the form 
\begin{equation}
\dot{x}(t)=f\left(x(t)\right)\label{eq:dynsys-1}
\end{equation}
with $x\left(t\right)\in\mathbb{R}^{n}$ and $f:\mathbb{R}^{n}\to\mathbb{R}^{n}$,
and assume that there exists a $\tau$ periodic solution $x^{*}\left(t\right)=x^{*}\left(t+\tau\right)$
of \eqref{eq:dynsys-1}. With this periodic orbit we associate a (principal)
\emph{fundamental matrix} $\Phi\left(t\right)$ which fulfills the
matrix equation 
\begin{equation}
\dot{\Phi}\left(t\right)=Df\left(x^{*}\left(t\right)\right)\Phi\left(t\right);\qquad\Phi\left(0\right)=\mathbb{I},\label{eq:Phidef-1}
\end{equation}
where $Df\left(x^{*}\left(t\right)\right)$ denotes the Jacobian of
$f$ evaluated at the point $x^{*}\left(t\right)$ along the periodic
orbit. For $t=\tau$ the fundamental matrix $\Phi\left(\tau\right)$
is often called \emph{monodromy matrix,} and the generalised eigenvalues
$\mu_{1},\ldots,\mu_{n}$ of $\Phi\left(\tau\right)$ are known as
\emph{Floquet multipliers} (or characteristic multipliers) of the
periodic orbit $x^{*}\left(t\right)$. Taking the time derivative
of \eqref{eq:dynsys-1} and taking into account that the function
$f$ does not explicitly depend on time, we observe that 
\[
\frac{d}{dt}\dot{x}^{*}\left(t\right)=Df\left(x^{*}\left(t\right)\right)\dot{x}^{*}\left(t\right).
\]
Comparing with \eqref{eq:Phidef-1} we can therefore identify $\dot{x}^{*}\left(t\right)=\Phi\left(t\right)\dot{x}^{*}\left(0\right)$
and in particular note that 
\[
\Phi\left(\tau\right)\dot{x}^{*}\left(0\right)=\dot{x}^{*}\left(\tau\right)=\dot{x}^{*}\left(0\right).
\]
It therefore follows that the\emph{ }monodromy matrix $\Phi\left(\tau\right)$
for a periodic orbit in an autonomous system has at least one eigenvalue
equal to one, and we choose in the following $\mu_{1}=1$. As $\Phi\left(\tau\right)$
is a real matrix, the set of the remaining Floquet multipliers $\left\{ \mu_{2},\ldots,\mu_{n}\right\} $
is composed of either real numbers, or complex conjugate pairs of
complex numbers. The significance of the Floquet multipliers lies
in the fact that they allow us to characterise the periodic orbit
in question. For example, if there exists at least one Floquet multiplier
$\mu_{k}$ such that $\left|\mu_{k}\right|>1$, then the periodic
orbit is unstable. If there exists at least one Floquet multiplier
$\mu_{k}$ with $k>1$ such that $\left|\mu_{k}\right|=1$, then the
orbit is called \emph{non-hyperbolic}. In the following we will assume
that the the periodic orbit $x^{*}\left(t\right)$ is hyperbolic,
i.e. none of the Floquet multipliers other than $\mu_{1}$ is located
on the unit circle. We also define the symbol $m$ to denote the number
of Floquet multipliers which are real and larger than one. For $m>0$
the periodic orbit is unstable, however the converse is not true. 

Following the ideas of \citep{SOC94} we now introduce an extended
time-delayed feedback term by modifying the system \eqref{eq:dynsys-1}
as follows

\begin{equation}
\begin{aligned}\dot{x}(t) & =f\left(x(t)\right)+Ky\left(t-0\right)\\
y\left(t\right) & =x(t-\tau)-x(t)+Ry\left(t-\tau\right)
\end{aligned}
\label{eq:ETDAS}
\end{equation}
where $y\left(t\right)\in\mathbb{R}^{n}$, $\tau$ is a positive parameter
for the delay time. Both, the control matrix $K$ and the memory matrix
$R$ are $n\times n$ matrices, where $R$ fulfills the condition
\[
\lim_{k\to\infty}R^{k}=0.
\]
The solution of \eqref{eq:ETDAS} for $t>0$ requires the knowledge
of the two (left-continuous) functions $x\left(t\right)$ and $y\left(t\right)$
on the interval $t\in\left(-\tau,0\right]$. Note that formally the
system \eqref{eq:ETDAS} is not a delay \emph{differential} equation,
since no time derivative of $y\left(t\right)$ appears. For $R=0$
we recover the traditional Pyragas control scheme. Instead of \eqref{eq:ETDAS}
many other essentially equivalent formulations of EDFC are used, for
example 
\[
\dot{x}(t)=f\left(x(t)\right)+K\left\{ \sum_{j=0}^{\infty}R^{j}\left[x\left(t-\left(j+1\right)\tau\right)-x\left(t-j\tau\right)\right]\right\} .
\]

The system \eqref{eq:ETDAS} possesses the obvious $\tau$ periodic
solution

\begin{equation}
\begin{aligned}x_{\tau}\left(t\right) & =x^{*}\left(t\right)\\
y_{\tau}\left(t\right) & =0,
\end{aligned}
\label{eq:soltau}
\end{equation}
however the stability of this solution may have been affected by the
control scheme. In the following we would like to gain some insight
into the stability properties of this solution. 

Before we formulate our main theorem, we slightly change the time-delay
parameter $\tau$ appearing in \eqref{eq:ETDAS} to a new value $\hat{\tau}$
as follows:
\begin{equation}
\begin{aligned}\dot{x}(t) & =f\left(x(t)\right)+Ky\left(t-0\right)\\
y\left(t\right) & =x(t-\hat{\tau})-x(t)+Ry\left(t-\hat{\tau}\right)
\end{aligned}
\label{eq:ETDAS-1}
\end{equation}
 If $\hat{\tau}$ is sufficiently close to $\tau$ we can assume that
there exists a periodic solution $\left(x_{\tilde{\tau}},y_{\tilde{\tau}}\right)$
for \eqref{eq:ETDAS-1} which is close to the original solution \eqref{eq:soltau}.
However, we expect that the period of this new solution will be in
general different from both, $\tau$ and $\hat{\tau}$, and we denote
this period by the new symbol $\tilde{\tau}.$ Continuity requires
that $\lim_{\hat{\tau}\to\tau}\tilde{\tau}\left(\hat{\tau}\right)=$$\tau$.
We can now formulate our main theorem.

\textbf{Theorem: }Let $x^{*}\left(t\right)$ be a $\tau$-periodic
orbit of \eqref{eq:dynsys-1} with $m$ real Floquet multipliers greater
than unity and let $\tilde{\tau}\left(\hat{\tau}\right)$ be the period
of the induced periodic orbit of \eqref{eq:ETDAS-1} with $\lim_{\hat{\tau}\to\tau}\tilde{\tau}\left(\hat{\tau}\right)=$$\tau$.
Then the orbit $x^{*}\left(t\right)$ is an unstable solution of \eqref{eq:ETDAS}
if the condition
\begin{equation}
\left(-1\right)^{m}\lim_{\hat{\tau}\to\tau}\frac{\hat{\tau}-\tau}{\hat{\tau}-\tilde{\tau}\left(\hat{\tau}\right)}<0,\label{eq:condition-1}
\end{equation}
is fulfilled.

\section{Proof of the theorem\label{sec:Proof-of-the}}

In the proof we follow the techniques developed in \citep{HOO12,NAK97,NAK98},
but we aim to keep the following presentation as self-contained as
possible. We first formulate a lemma, which allows us to connect the
linear stability of system \eqref{eq:ETDAS} with the solution of
a time dependent linear equation.

\textbf{Lemma 1:} If the equation 
\begin{equation}
\frac{d}{dt}\left[\delta x\left(t\right)\right]=\left\{ Df\left(x^{*}\left(t\right)\right)+\left(\nu^{-1}-1\right)K\left[1-R\nu^{-1}\right]^{-1}\right\} \delta x\left(t\right),\label{eq:lin1}
\end{equation}
possesses a solution of the form 
\begin{equation}
\delta x\left(t\right)=\nu\delta x\left(t-\tau\right)\label{eq:numult}
\end{equation}
 for real $\nu>1$, then the periodic orbit $x^{*}\left(t\right)$
is an unstable solution of \eqref{eq:ETDAS}.

Proof: First note that because of the requirement $\lim_{k\to\infty}R^{k}=0$,
the eigenvalues of $R\nu^{-1}$ are contained in the unit circle of
the complex plane, and the matrix $\left[1-R\nu^{-1}\right]$ is indeed
non-singular for all $\nu\geq1$. It is then straightforward to check
that the ansatz 
\begin{eqnarray*}
x\left(t\right) & = & x^{*}\left(t\right)+\delta x\left(t\right)\\
y\left(t\right) & = & \left[1-R\nu^{-1}\right]^{-1}\left(\nu^{-1}-1\right)\delta x(t)
\end{eqnarray*}
fulfills \eqref{eq:ETDAS} to first order in $\delta x\left(t\right)$.
Because of \eqref{eq:numult} we have therefore explicitly constructed
an exponentially growing linear perturbation of the solution \eqref{eq:soltau},
which implies that $x^{*}\left(t\right)$ is an unstable solution
of \eqref{eq:ETDAS}. This concludes the proof of Lemma 1.

We now have to show that under the conditions of the theorem there
exists a $\nu>1$ such that the linear equation \eqref{eq:lin1} allows
for a solution fulfilling \eqref{eq:numult}. We first introduce the
fundamental matrix for \eqref{eq:lin1} via 
\begin{eqnarray}
\dot{\Psi}_{\nu}\left(t\right) & = & \left\{ Df\left(x^{*}\left(t\right)\right)+\left(\nu^{-1}-1\right)K\left[1-R\nu^{-1}\right]^{-1}\right\} \Psi_{\nu}\left(t\right),\label{eq:Psidef-1}\\
\Psi_{\nu}\left(0\right) & = & \mathbb{I}
\end{eqnarray}
If we now find a $\nu>1$ such that 
\begin{equation}
\det\left(\nu\mathbb{I}-\Psi_{\nu}\left(\tau\right)\right)=0,\label{eq:detcond}
\end{equation}
then there exists a $\delta x_{0}\in\mathbb{R}^{n}$ which fulfills
$\Psi_{\nu}\left(\tau\right)\delta x_{0}=\nu\delta x_{0}$. Using
$\delta x\left(t\right)=\Psi_{\nu}\left(t\right)\delta x_{0}$ we
can then construct a solution of \eqref{eq:lin1} which is of the
form required by \eqref{eq:numult}. This motivates the introduction
of the function \citep{NAK97} 
\begin{equation}
F\left(\nu\right)=\det\left(\nu\mathbb{I}-\Psi_{\nu}\left(\tau\right)\right)\label{eq:detF-1}
\end{equation}
and from the above discussion we can conclude that the proof of our
theorem is complete, if we are able to show that there exists a $\nu_{c}>1$
such that $F\left(\nu_{c}\right)=0.$ For $\nu=1$ we see from \eqref{eq:Psidef-1}
that $\Psi_{1}\left(t\right)=\Phi\left(t\right)$. One of the eigenvalues
of $\Phi\left(\tau\right)$ is however equal to unity, and we therefore
find that $F\left(1\right)=0.$ 

For the further discussion of $F\left(\nu\right)$ we write the fundamental
matrix in the form 
\begin{equation}
\Psi_{\nu}\left(t\right)=\Phi\left(t\right)\left[\mathbb{I}+\left(\nu^{-1}-1\right)\int_{0}^{t}\Phi^{-1}\left(u\right)K\left[1-R\nu^{-1}\right]^{-1}\Psi_{\nu}\left(u\right)du\right],\label{eq:integralPsi}
\end{equation}
which can be easily checked by direct differentiation. All the terms
appearing in this expression for $\Psi_{\nu}\left(t\right)$ remain
finite for $\nu\to\infty,$ and therefore \eqref{eq:detF-1} yields
$\lim_{\nu\to\infty}\frac{F\left(\nu\right)}{\nu^{n}}=1$, or in other
words, $F\left(\nu\right)$ diverges as $\nu^{n}$ for large $\nu.$
We can therefore summarise the discussion in the last two paragraphs
in the following lemma.

\textbf{Lemma 2}: If for a given periodic orbit $x^{*}\left(t\right)$
the condition 
\begin{equation}
\left.\frac{\partial F\left(\nu\right)}{\partial\nu}\right|_{\nu=1}=F'\left(1\right)<0\label{eq:cond_deriv}
\end{equation}
holds, then the orbit is an unstable solution of \eqref{eq:ETDAS}.

Proof: As $F\left(\nu\right)$ is continuous, $F\left(1\right)=0$
and $F'\left(1\right)<0$ there exists a $\nu_{n}>1$ such that $F\left(\nu_{n}\right)<0$.
However as $F\left(\nu\right)$ diverges as $\nu^{n}$ for large $\nu,$
it follows by the intermediate value theorem that there exists $\nu_{c}>1$
with $F\left(\nu_{c}\right)=0$. For this $\nu_{c}$ we can then construct
the function $\delta x\left(t\right)$ required for Lemma 1 and the
orbit is unstable. Therefore Lemma 2 is proved. 

In view of Lemma 2 we now need to study the conditions for which $F'\left(1\right)<0$
holds. Before we proceed it is now useful to introduce the matrix
$W$, which diagonalizes $\Phi\left(\tau\right)$, i.e.
\[
\left(\begin{array}{ccccc}
1 & 0 & 0 & \cdots & 0\\
0 & \mu_{2} & * & \ddots & \vdots\\
\vdots & \ddots & \ddots & \ddots & 0\\
\vdots &  & \ddots & \ddots & *\\
0 & \cdots & \cdots & 0 & \mu_{n}
\end{array}\right)=W^{-1}\Phi\left(\tau\right)W.
\]
 Here the $*$ denotes either $0$ or $1$, depending on the Jordan
block associated with a particular eigenvalue. This allows us to formulate 

\textbf{Lemma 3: }A periodic orbit with $m$ Floquet multipliers greater
than unity is an unstable solution of the EDFC scheme \eqref{eq:ETDAS}
if the condition 
\begin{equation}
\left(-1\right)^{m}\left(1+\int_{0}^{\tau}\left(W^{-1}\Phi\left(t\right)^{-1}K\left[1-R\right]^{-1}\Phi\left(t\right)W\right)_{11}dt\right)<0\label{eq:cond_analytic}
\end{equation}
holds. 

Proof: The idea of the proof is to show that \eqref{eq:cond_analytic}
implies \eqref{eq:cond_deriv} and then use Lemma 2. To calculate
the sign of $F'\left(1\right)=\lim_{\epsilon\to0}F\left(1+\epsilon\right)/\epsilon$
we expand $F\left(1+\epsilon\right)$ to first order in $\epsilon$.
We obtain 
\begin{eqnarray}
F\left(1+\epsilon\right) & = & \det\left[\left(1+\epsilon\right)\mathbb{I}-\Psi_{1+\epsilon}\left(\tau\right)\right]=\det\left[W^{-1}\left(\left(1+\epsilon\right)\mathbb{I}-\Psi_{1+\epsilon}\left(\tau\right)\right)W\right]\nonumber \\
 & = & \det\left[M^{0}+\epsilon M^{1}\right]+O\left(\epsilon^{2}\right)\label{eq:det}
\end{eqnarray}
where the matrices $M^{0}$ and $M^{1}$ collect the terms in zeroth
and first order of $\epsilon$. We obtain
\[
\begin{gathered}W^{-1}\left(\left(1+\epsilon\right)\mathbb{I}-\Psi_{1+\epsilon}\left(t\right)\right)W\\
=\left(1+\epsilon\right)\mathbb{I}-W^{-1}\Phi\left(\tau\right)\left[\mathbb{I}+\left(\frac{1}{1+\epsilon}-1\right)\int_{0}^{\tau}\Phi^{-1}\left(u\right)K\left[1-R\frac{1}{1+\epsilon}\right]^{-1}\Psi_{1+\epsilon}\left(u\right)du\right]W\\
=\left(1+\epsilon\right)\mathbb{I}-W^{-1}\Phi\left(\tau\right)W+\epsilon\int_{0}^{\tau}W^{-1}\Phi^{-1}\left(u\right)K\left[1-R\right]^{-1}\Phi\left(u\right)duW+O\left(\epsilon^{2}\right)\\
=\mathbb{I}-W^{-1}\Phi\left(\tau\right)W+\epsilon\left[\mathbb{I}+\int_{0}^{\tau}W^{-1}\Phi^{-1}\left(u\right)K\left[1-R\right]^{-1}\Phi\left(u\right)Wdu\right]+O\left(\epsilon^{2}\right)
\end{gathered}
\]
and therefore 
\[
M^{0}=W^{-1}\left(\mathbb{I}-\Phi\left(t\right)\right)W=\left(\begin{array}{ccccc}
0 & 0 & 0 & \cdots & 0\\
0 & 1-\mu_{2} & * & \ddots & \vdots\\
\vdots & 0 & \ddots & \ddots & 0\\
\vdots &  & \ddots & \ddots & *\\
0 & \cdots & \cdots & 0 & 1-\mu_{n}
\end{array}\right)
\]
 and 
\[
M^{1}=\mathbb{I}+\int_{0}^{\tau}W^{-1}\Phi^{-1}\left(u\right)K\left[1-R\right]^{-1}\Phi\left(u\right)Wdu
\]
 Because of the special form of $M^{0}$, only one term contributes
to the determinant in \eqref{eq:det} in first order of $\epsilon$,
\begin{eqnarray*}
F\left(1+\epsilon\right) & = & \det\left[M^{0}+\epsilon M^{1}\right]+O\left(\epsilon^{2}\right)=\epsilon M_{11}^{1}\prod_{k=2}^{n}\left(1-\mu_{k}\right)+O\left(\epsilon^{2}\right)
\end{eqnarray*}
 For the sign of $F'\left(1\right)$ we therefore conclude 
\begin{equation}
\mbox{sgn}\left(F'\left(1\right)\right)=\mbox{sgn}\left(M_{11}^{1}\prod_{k=2}^{n}\left(1-\mu_{k}\right)\right)=\mbox{sgn}\left(M_{11}^{1}\left(-1\right)^{m}\right)\label{eq:signcond}
\end{equation}
where in the last step we used that every real Floquet multiplier
greater than unity contributes a minus sign to the product. Comparing
with \eqref{eq:cond_analytic} we observe that if condition \eqref{eq:cond_analytic}
is fulfilled then condition \eqref{eq:cond_deriv} is also fulfilled
because of \eqref{eq:signcond}. Therefore Lemma 3 now follows by
evoking Lemma 2.

As a final step in the proof of the main theorem, it now remains to
be shown that \eqref{eq:condition-1} implies condition \eqref{eq:cond_analytic}.
We need to establish, how the detuning between the period $\tau$
of the uncontrolled orbit and the delay time $\hat{\tau}$ of the
EDFC scheme \eqref{eq:ETDAS-1} influence the period $\tilde{\tau}$
of the orbit which is induced by this detuning. For the Pyragas control
this problem was solved by Just et al. in \citep{JUS98} and for the
case of EDFC by Novi\v{c}enko and Pyragas in \citep{NOV12a}. In our
notation the result of equation (28) from \citep{NOV12a} can be written
in the form 
\[
\tilde{\tau}=\tau+\left(\hat{\tau}-\tau\right)\frac{M_{11}^{1}-1}{M_{11}^{1}}+O\left[\left(\hat{\tau}-\tau\right)^{2}\right]
\]
or 
\begin{eqnarray*}
M_{11}^{1} & = & \frac{\hat{\tau}-\tau}{\hat{\tau}-\tilde{\tau}}
\end{eqnarray*}
It therefore follows that the conditions \eqref{eq:cond_analytic}
in Lemma 3 and the condition \eqref{eq:condition-1} in the main theorem
are equivalent, and this concludes the proof of the theorem.

\section{Discussion\label{sec:Discussion-and-conclusion}}

Our main theorem provides a limitation on the applicability of EDFC,
which involves the number of real Floquet multipliers ($m$) and a
combination of the period of the uncontrolled system $\left(\tau\right)$,
a detuned delay time $\left(\hat{\tau}\right)$ and the resulting
period of the induced orbit $\left(\tilde{\tau}\right)$. Let us now
have a closer look at the condition \eqref{eq:condition-1} appearing
in the theorem. We see that $m$ only appears in the form $\left(-1\right)^{m}$,
which is negative for odd $m$ and positive otherwise. The second
factor involves a ratio $r$ of the form 
\begin{equation}
r=\frac{\hat{\tau}-\tau}{\hat{\tau}-\tilde{\tau}},\label{eq:ratio}
\end{equation}
and we are asked to evaluate the sign of this ratio in the limit where
$\hat{\tau}$ goes to $\tau$. If this ratio was always positive,
then our theorem would simply reduce to the statement of the old odd-number
limitation, i.e. orbits with odd number $m$ cannot be controlled.
Our theorem now modifies this statement as follows: orbits with odd
$m$ can be stabilised, but only if the control terms are implemented
in such a way that $r$ is negative. We stress that our theorem only
gives a necessary condition for control, i.e. a violation of condition
\eqref{eq:condition-1} will not guarantee that an orbit will be stabilised.
Intuitively the case of negative $r$ seems to be unusual, as it implies
that if we slightly increase the delay time $\hat{\tau}$ to a value
larger than $\tau$, then the period $\tilde{\tau}$ of the induced
orbit needs to be even larger than $\hat{\tau}$ itself. This intuitively
strange situation might be one of the reasons, why the violation of
the original odd-number limitation in the autonomous case was not
observed earlier. In the case of the counter example given in \citep{FIE07},
$r$ is indeed negative, as was shown in \citep{HOO12}.

We also remark that Lemma 3 provides useful insight in its own right.
We can slightly rewrite the condition \eqref{eq:cond_analytic} as
follows 
\[
\left(-1\right)^{m}\left(1+\int_{0}^{\tau}\left(z^{T}\left(t\right)K\left[1-R\right]^{-1}\dot{x}^{*}\left(t\right)\right)dt\right)<0
\]
 Here $z^{T}\left(t\right)$ is the first row of the matrix $W^{-1}\Phi\left(t\right)^{-1}$,
and is also known as the dual vector of the zero mode. If we now write
the control matrix with a scalar prefactor $k$ in the form 
\[
K=kK_{0}
\]
 and introduce $\kappa$ via 
\[
-\kappa^{-1}=\int_{0}^{\tau}\left(z^{T}\left(t\right)K_{0}\left[1-R\right]^{-1}\dot{x}^{*}\left(t\right)\right).
\]
Then the condition \eqref{eq:cond_analytic} can be written in the
form 
\begin{equation}
\left(-1\right)^{m}\left(1-\frac{k}{\kappa}\right)<0.\label{eq:kappacond}
\end{equation}
The period of the induced orbit $\tilde{\tau}\left(\hat{\tau},k\right)$
now depends on the scalar coupling strength $k$ and the delay time
$\hat{\tau}$ in the following way \citep{JUS98,NOV12a}, 
\begin{equation}
\tilde{\tau}\left(\hat{\tau},k\right)=\tau+\frac{k}{k-\kappa}\left(\hat{\tau}-\tau\right)+O\left[\left(\hat{\tau}-\tau\right)^{2}\right].\label{eq:tautilde}
\end{equation}
 This allows us to conveniently determine the parameter $\kappa$
through 
\begin{equation}
\kappa=\lim_{k\to0}\lim_{\hat{\tau}\to\tau}k\frac{\tilde{\tau}\left(\hat{\tau},k\right)-\hat{\tau}}{\tilde{\tau}\left(\hat{\tau},k\right)-\tau}\label{eq:kappacalc}
\end{equation}
If we focus our interest again to the case, where the stabilisation
for odd $m$ is possible, we can conclude from \eqref{eq:kappacond}
that this requires a positive $\kappa$. From \eqref{eq:kappacalc}
we can obtain the value and in particular the sign of $\kappa$ from
a ``measurement'' of $\tilde{\tau}$ in the low coupling regime.
From \eqref{eq:tautilde} we observe that for positive $\kappa$ the
period of the induced orbit $\tilde{\tau}$ is always outside the
interval $\left[\tau,\hat{\tau}\right]$ for all positive values of
$k$, and $\tilde{\tau}$ diverges at the point where \eqref{eq:kappacond}
is first violated.

\section*{Acknowledgements}

We thank B. Fiedler, V. Flunkert, P. H{\"o}vel, W. Just, K. Pyragas and
E. Sch{\"o}ll for fruitful discussions. This work was supported by Science
Foundation Ireland under Grant Number 09/SIRG/I1615.

%

\end{document}